\documentclass[twocolumn,preprint,5p]{elsarticle}

\usepackage{natbib,graphicx,lineno,txfonts,color,slashbox,subfigure}

\newcommand{\etal}{\hbox{et al.} }
\newcommand{\ADD}{\textcolor{black}}
\newcommand{\REMOVE}[1]{}

\journal{Astroparticle Physics}
\begin{document}
\begin{frontmatter}
\title{Gamma-Hadron Separation in Very-High-Energy $\gamma$-ray astronomy using a multivariate analysis method}

\author[mpi]{S. Ohm\corref{cor}}
\ead{stefan.ohm@mpi-hd.mpg.de}
\cortext[cor]{Corresponding author.}
\author[mpi]{C. van Eldik}
\ead{christopher.van.eldik@mpi-hd.mpg.de}
\author[mpi]{K. Egberts}
\ead{kathrin.egberts@mpi-hd.mpg.de}

\address[mpi]{Max-Planck-Institut f\"ur Kernphysik, Heidelberg, Germany}
\begin{abstract}
In recent years, Imaging Atmospheric Cherenkov Telescopes (IACTs) have discovered a rich diversity of very high 
energy (VHE, $> 100$~GeV) $\gamma$-ray emitters in the sky. These instruments image Cherenkov light emitted by 
$\gamma$-ray induced particle cascades in the atmosphere. Background from the much more numerous cosmic-ray cascades 
is efficiently reduced by considering the shape of the shower images, and the capability to reduce this background 
is one of the key aspects that determine the sensitivity of a IACT. In this work we apply a tree classification method 
to data from the High Energy Stereoscopic System (H.E.S.S.). We show the stability of the method and its capabilities 
to yield an improved background reduction compared to the H.E.S.S. Standard Analysis.

\end{abstract}
\begin{keyword}
classification, separation, decision tree, $\gamma$-ray astronomy, Cherenkov technique
\end{keyword}
\end{frontmatter}

\section{Introduction}
\label{sec:Intro}

In the last years ground-based Imaging Atmospheric Cherenkov Telescopes (IACTs) opened a previously inaccessible window 
for the study of astrophysical sources of $\gamma$ radiation in the VHE regime. The detection of more than 50 galactic 
VHE $\gamma$-ray emitters during the galactic plane scan performed by the H.E.S.S. collaboration between 2004 and 2007 
\cite{Aharonian2005,Aharonian2006,Hoppe2007} decupled the number of known VHE $\gamma$-ray sources and hence established 
a new field in astronomy. 

The earth's atmosphere is opaque to VHE photons, which initiate electromagnetic particle cascades (Extensive Air 
Showers, EAS) in the atmosphere. The highly relativistic charged particles in the cascade emit Cherenkov light which can be 
imaged via a large mirror onto a fine-grained camera. From the shower image one can reconstruct the arrival direction 
of the primary $\gamma$-ray and calculate its energy using the number of collected Cherenkov photons and the directional 
information.

One of the big advantages of IACTs is their enormous effective detector area. Modern instruments reach 
$\sim10^5~\mathrm{m^2}$ which is five orders of magnitude larger than what is typically achieved with 
satellite-based instruments like EGRET or Fermi LAT. While the latter benefit from quasi background free observations, 
Cherenkov telescopes have to deal with a vast number of hadronic cosmic-ray background events. The capability to suppress these 
against the $\gamma$-rays associated with astrophysical sources is one of the key aspects that determines the sensitivity 
of IACTs.

To increase the sensitivity of ground-based VHE $\gamma$-ray telescope systems beyond what is obtained with 
state-of-the-art instruments like H.E.S.S. \cite{Hinton2004}, MAGIC \cite{Lorenz2004}, VERITAS \cite{Weekes2002} or 
CANGAROO-III \cite{Kubo2004} larger arrays are needed, as e.g. studied by the CTA \cite{CTA2007} and AGIS 
\cite{Fegan2008} consortia.

Still, for the existing instruments, increased background reduction can improve the sensitivity considerably. With respect 
to the classical - robust but less sensitive - {\it Hillas approach} 
\cite{Hillas1985}, which parametrises the 2-dimensional elliptical shape of the recorded images for reconstruction and 
selection of $\gamma$-ray like events, sensitivity can be increased by e.g. analysis methods which compare the detected 
images with a 3-dimensional photosphere model of the EAS (e.g. {\it 3D Model analysis}, introduced by Lemoine-Goumard \etal 
\cite{Marianne2006}). Furthermore, the applicability of 
multivariate analysis techniques \footnote{These techniques combine several shower parameters into one number which 
gives the likeness of an event with a $\gamma$-ray or a cosmic ray.} like Random Forests \cite{Breiman2001} in 
ground-based VHE astronomy has recently been demonstrated \cite{Albert2008a,Egberts2008,Bock2004}.

In this paper we follow the latter approach and discuss the application of the {\it Boosted Decision Trees} (BDT) 
method, provided by the TMVA package \cite{tmva2007}, to data obtained by the H.E.S.S. experiment. The stability of 
the technique and its capabilities to improve $\gamma$/hadron separation compared to the H.E.S.S. Standard Analysis are 
demonstrated. After a brief description of the method (Chapter \ref{section:BDT}) and an introduction of the training 
and evaluation of the BDT method with events recorded by the H.E.S.S. experiment in Chapter \ref{section:application}, 
the application of BDT to H.E.S.S. data is discussed (Chapter \ref{section:comparison}). Finally, performance and 
sensitivity evaluated using Monte-Carlo simulations and background data are presented (Chapter \ref{section:tests}). 

\section{Classification using Boosted Decision Trees} \label{section:BDT}

Machine learning algorithms like Neural Networks (NNs), Likelihood Estimators or Fisher discriminants are basically 
extensions of simple cut-based analysis techniques to multivariate algorithms. They are widely used in natural sciences 
for classification of events of different type which are described by a set of input parameters. Beyond the 
aforementioned techniques the MiniBooNE \cite{Roe2005,Yang2005} and D0 \cite{Benitez2008} collaborations recently 
utilised the BDT method for particle identification in high energy physics, and Bailey \etal \cite{Bailey2007} 
used it for supernova searches in optical astronomy.

One of the main advantages of NNs and BDT compared to Likelihood classifiers or Fisher discriminants is the consideration 
of nonlinear correlations between input parameters. Furthermore, the BDT method effectively ignores parameters without 
separation power whereas NNs could suffer from those, resulting in a degraded separation. 

\subsection{Basics of the Decision Tree Algorithm} \label{subsection:BDTmethod}

\begin{figure}
\begin{center}
  \includegraphics[scale=0.2]{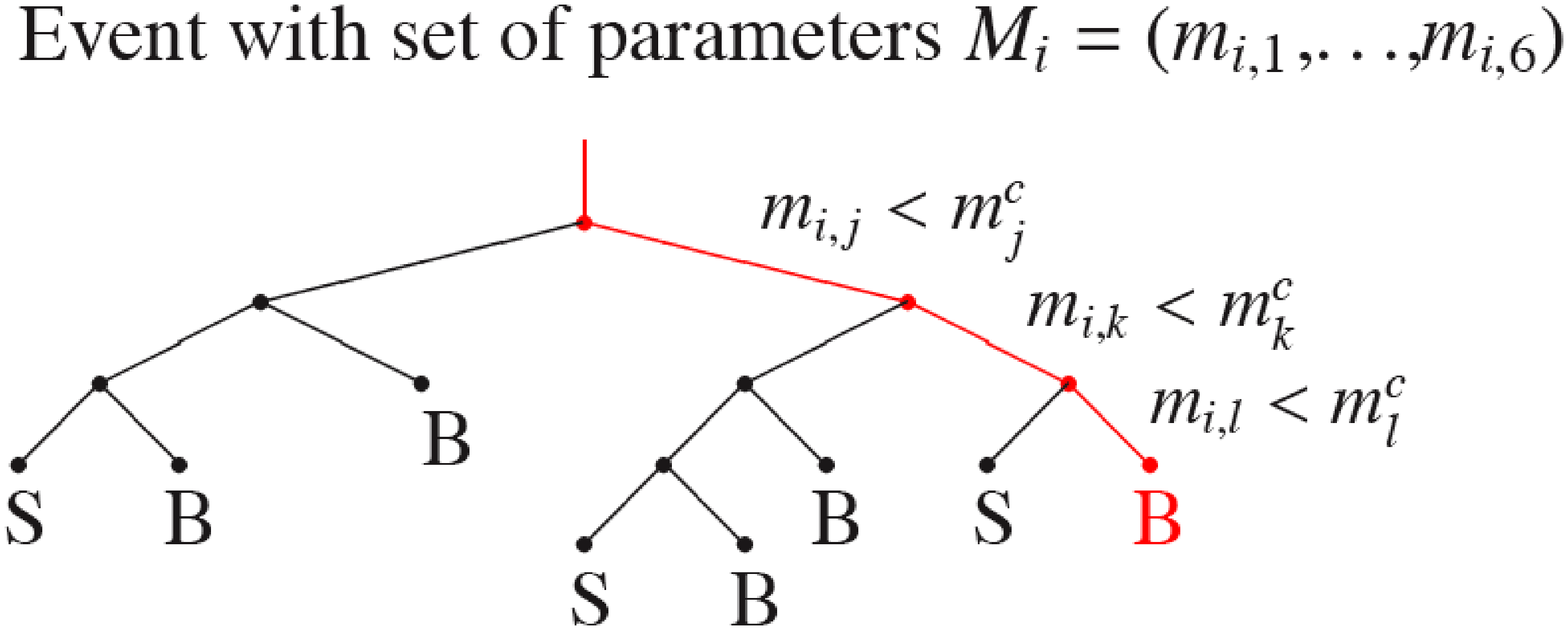}
  \caption{Sketch of a decision tree. An event, described by a parameter set, $M_i$ = ($m_{i,1}$,$\ldots$,$m_{i,6}$), 
undergoes at each node a binary split criterion (passed or failed) on one of its parameters until it ends up in a 
{\it leaf}. This leaf marks it as signal (S) or background (B).}
  \label{fig:method}
\end{center}
\end{figure}

Decision trees \cite{Breiman1984,Bowser1993} can be represented by a two dimensional structure like the one sketched 
in Fig. \ref{fig:method}. By applying, at each branching, a binary split criterion (passed or failed) on one of the 
characterising input parameters they classify events of unknown type as signal-like or background-like. The determination 
of these criteria is also referred to as {\it training} of a decision tree, and is performed with a training set 
consisting of events of known type. To circumvent a drawback of single decision trees, namely the instability against 
statistical fluctuations in the training event set, one extends the single decision tree to a {\it forest} of decision 
trees, which differ in the binary split criteria. A weighted mean vote of the classification of all single trees in the 
forest stabilises the response of the classifier and improves its performance. This vote is 
the output of the BDT and describes the signal- or background-likeliness of an event. In this work it is referred to as 
the $\zeta$ variable. The forest of trees is obtained by a process called ``boosting'', starting from an initial single tree.

\subsection{The training procedure for a single tree} \label{subsection:trainingbasic}

The training or building of a decision tree is the process that finds the appropriate splitting criterion for each node 
using a training sample, $S$, of events of known type. The training sample is composed of a signal training sample, $S_1$, 
and a background training sample, $S_2$, which consist of $N_1$ signal and $N_2$ background events, respectively. 
Each event in the training set is characterised by a weighting factor $\omega_i$ and a set of input parameters, $M_i$. 
To build a decision tree from such a training sample the following steps are performed:
\begin{itemize}
\item  The training samples are normalised in such a way that all signal events have the same weight, 
$\omega_{i}(S_1) = 1/N_1$, and all background events have the same weight $\omega_{i}(S_2) = 1/N_2$.
\item Tree building starts at the root node (top node in Fig. \ref{fig:method}), where one finds the variable and split 
value that provides the best separation of signal and background events. According to this splitting criterion, $S$ is 
divided into two subsets of events that either pass or fail this criterion. Each subset is fed into a child node where 
again the cut parameter which separates best signal and background events is determined. 
\item This procedure is applied recursively until further splitting would not increase the separation, or a preassigned 
minimum number of events is reached \footnote{This avoids overtraining due to statistically insignificant leaves.}. 
According to the majority of signal and background events, the last-grown nodes (which are called leaves) are assigned 
signal (S) or background (B) type, respectively (see Fig. \ref{fig:method}).
\end{itemize}

\subsection{Boosting} \label{boosting}
Single decision trees are sensitive to statistical fluctuations in the training sample, hence a boosting procedure is applied 
which results in a forest of trees and thus increases the stability of the method. In this procedure, events that got 
misclassified in the building of the previous tree are multiplied with a {\it boost weight}, $\alpha$, thereby getting a 
higher weight in the training of the next tree. Hence, the boosting is applied to all trees except for the first one.
This method is known as {\it AdaBoost} or adaptive boost 
\cite{AdaBoost1996}. $\alpha$ is calculated from the fraction of misclassified events in all leaves, $err$: 

\begin{equation}
\alpha = \frac{1 - err}{err}
\end{equation}
After having applied $\alpha$ to each misclassified event, renormalisation of the training samples retains the sum of 
weights of all events in a decision tree constant.

\subsection{BDT settings} \label{BDTparameters}

In this work we use the BDT method provided by the TMVA package. The decision tree settings are mostly default values, 
which have been optimised and tested by the TMVA developers. These parameters guarantee a fast training procedure and a 
stable response of the classifier and are marked with a * in the following.
\begin{itemize}
\item The number of trees was chosen to be 200*, which is a compromise between separation performance and processing power.
Varying this value in a broad range does not significantly change the presented results.
\item The {\it Gini Index*} was used as separation type. It calculates the inequality between signal and background 
distributions for each value to find the best cut. \ADD{Other separation types were tested and found to achieve similar 
results.}
\item Splitting was stopped when the number of events in a node fell below \REMOVE{20*.} 
\ADD{($N_1$ + $N_2$) / (10 $\cdot N_{par}^2$)*, taking into account the training statistics and the number of 
training parameters. Typical numbers are between 100 and 1000 for the smallest and largest data set, respectively.}
\item The number of steps used to scan the parameter space for the best splitting criterion was increased from 20 to 
100 to adequately cover training parameters with a large range of values.
\end{itemize}

\section{Training and Evaluation of the BDT method} \label{section:application}

Having discussed the basic BDT-functioning and details of the growing procedure in the last chapter, this section deals 
with the training and evaluation of the BDT method. After an introduction of the training parameters used in this work, 
the properties of the signal- and background training sample are discussed. Finally, tests of the classifiers response 
are presented.

\subsection{Training parameters} \label{subsection:trainingparameters}

The recorded EAS images contain pixels which mainly store photons from the night sky background (NSB). They are 
removed in an image cleaning procedure \cite{Crab2006} for the further image analysis. Only pixels with an 
intensity of 5 p.e. and a neighbouring pixel with more than 10 p.e. (and vice versa) are kept, thereby just selecting 
pixels which contain Cherenkov photons originating from the EAS.

\begin{figure*}
\begin{center}
  \includegraphics[scale=0.7]{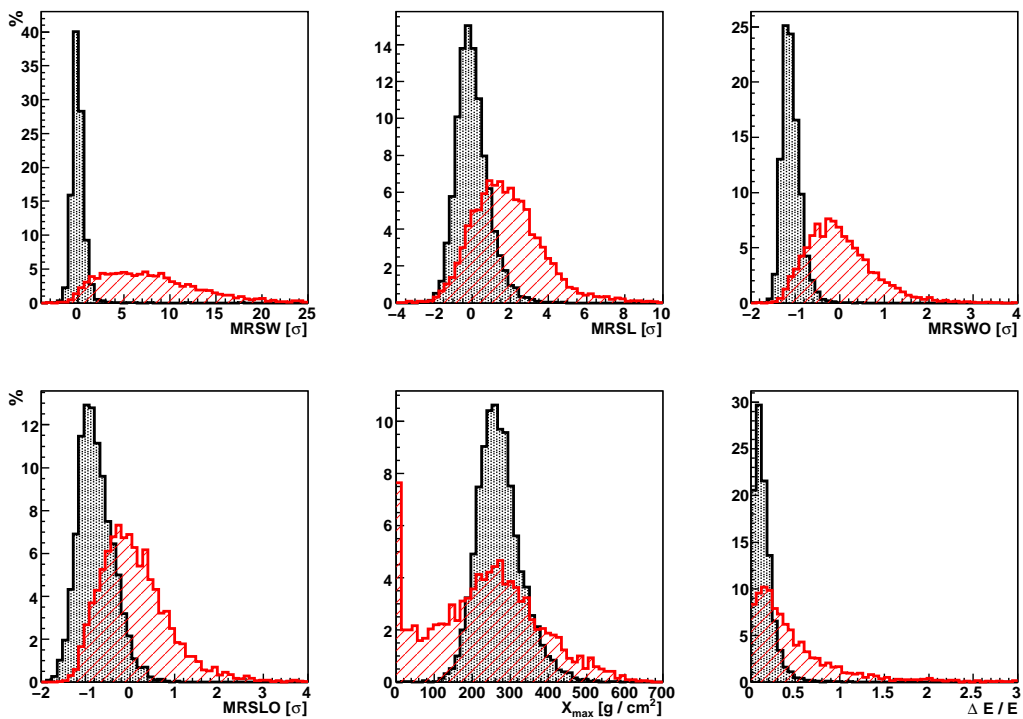}
  \caption{Distribution of the training variables with reconstructed energies between (0.5--1.0)~TeV in the zenith angle 
range (15--25)$^{\circ}$ for $\gamma$-rays (black) and cosmic-rays (red).}
  \label{fig:trainpar}
\end{center}
\end{figure*}

To classify the recorded air-shower events as of either signal- or background type, a set of training parameters has been 
derived using information from the EAS images. The training parameters are based on the {\it Hillas Parameters} 
\cite{Hillas1985} which are calculated using the second moments of the cleaned shower images. Of these, the width, length, 
and total intensity (also called {\it image size}) of the ellipse are used for classification. Compared to cosmic-ray induced 
showers, which in general exhibit a rather irregular shape, showers produced by $\gamma$-rays (or electrons) have an elliptical, 
quite regular structure. The Hillas Parameters inherently store information about the shape of the shower, and can therefore be 
used to discriminate between cosmic-ray and $\gamma$-ray primaries. Furthermore, an event recorded by multiple telescopes is 
better constrained. To be independent of the number of participating telescopes (hereafter called {\it multiplicity}), the 
Hillas Parameters of individual telescopes are averaged. The same is true for all the BDT training parameters, presented in 
the following:

\begin{itemize}
\item One type of training parameters is based on the {\it mean reduced scaled width} approach introduced by Aharonian \etal 
\cite{Crab2006}. For an image with a given size and reconstructed impact distance \footnote{The distance between the 
telescope and the impact point of the lengthened primary particle track on ground.} the mean expected width for a $\gamma$-ray 
$\langle W_i \rangle$ as obtained from $\gamma$-ray simulations is compared to the measured width $W_i$. The {\it Scaled Width} 
for telescope $i$ is then defined as $\mathrm{SCW}_i = (W_i - \langle W_i \rangle) / \sigma_i$, with $\sigma_i$ being the 
spread of the expected width. The mean reduced scaled width (MRSW) can then be calculated as the average SCW over all 
telescopes:

\begin{equation}
\label{eqn:mscw}
\mathrm{MRSW} = \frac{1}{\displaystyle\sum_{i\in N_{tel}}\omega_i} \cdot \displaystyle\sum_{i\in N_{tel}}
\left(\mathrm{SCW}_i \cdot \omega_i \right)\mbox{ ,}
\end{equation}
taking into account the accuracy of the $\gamma$-ray simulations by introducing a weighting factor $\omega_i$, defined as 
$\omega_i = \langle W_i \rangle ^2/\sigma_i^2$.

Similarly, the {\it mean reduced scaled length} (MRSL) is calculated. By comparing the measured width and length of the 
image with the prediction for an hadronic event \footnote{\ADD{These hadronic events are obtained in H.E.S.S. observations of 
sky regions without significant $\gamma$-ray contamination as cosmic-ray background (also referred to as {\it Off-Events})}}, 
two additional training parameters, the {\it mean reduced scaled width off} (MRSWO) and {\it mean reduced scaled length off} 
(MRSLO), are computed.

\item Another parameter addresses the different interaction lengths of photons and hadronic cosmic-rays in the atmosphere. 
It is expressed as the depth of the shower maximum X$_{\mathrm{max}}$ and reconstructed from the recorded shower images. 
Also this parameter is calculated as a weighted mean value over all participating telescopes. 
\item Because of their irregular structure, the energy of hadron-induced showers may be reconstructed differently for 
telescopes seeing the shower from different directions. The $\Delta$E / E parameter, calculated as the averaged spread 
in energy reconstruction between the triggered telescopes, adds additional separation power to the BDT classification.
\end{itemize}
For illustration, Fig. \ref{fig:trainpar} shows all training parameter distributions for events with zenith angles around 
20$^{\circ}$ and energies 0.5~TeV $\leq$ E $\leq$ 1.0~TeV. 

\subsection{Training sample} \label{subsection:trainingsample}
The training set used for building the BDT consists of Monte-Carlo simulations of $\gamma$-rays as signal events, and 
\REMOVE{events from H.E.S.S. observations of sky regions without significant $\gamma$-ray contamination as cosmic-ray 
background (also referred to as {\it Off-Events}).} \ADD{Off-Events as cosmic ray background.}
The $\gamma$-rays are simulated as resulting from a point source, at a fixed distance 
(offset) of 0.5$^{\circ}$ from the camera centre, and follow an energy distribution dN/dE $\sim$ E$^{-\Gamma}$ with index 
$\Gamma$ = 2.0. Since cosmic-rays reach the earth isotropically, the Off-Events are homogeneously distributed over the field 
of view of the camera. A cut on the minimum image size of 80 p.e. and the maximum distance between the centre-of-gravity 
($COG$) of the Hillas ellipse and the camera centre (to reduce effects of image truncation) was applied to the training sample. 
This is also referred to as {\it pre-selection} and used to exclude poorly reconstructed events from the training process.

\subsection{Training} \label{subsection:training}
The aim of a BDT classification is a stable $\gamma$/hadron separation over the whole dynamical range of the telescope 
system, which comprises the accessible energy range as well as the observational conditions (e.g. the zenith angle of the 
observation). Since the shower shape changes with the primary particle energy and its zenith angle, the distributions 
of \ADD{some of} the input parameters and consequently the response of the classifier changes. \ADD{As opposed to the mean 
reduced scaled parameters which by construction are independent of event energy and zenith angle, the depth of the shower 
maximum and the uncertainty in the energy estimation do depend on both these quantities.}

This characteristic requires a 
training of the BDT in energy- and zenith angle bands. The energy range accessible for H.E.S.S. (from $\sim$100~GeV 
to $\sim$100~TeV) was divided into six bands, based on the energy reconstructed assuming a $\gamma$-ray hypothesis, such that 
for each of seven zenith angle bands (from 0$^\circ$ to 60$^\circ$) the input parameter distributions do not change significantly, 
and a sufficient number of events for the training process was available. A summary of the training statistics in the energy- 
and zenith angle bands can be found in Table \ref{table:training}. The decreasing number of training events with increasing 
energy and/or zenith angle is a direct consequence of the energy spectra of the training sample and the increased energy threshold 
of the H.E.S.S. system at larger zenith angles. 

\begin{table*}
\begin{center}
  \begin{tabular}{|c|c|c|c|c|c|c|} \hline
    \multicolumn{1}{|c|}{\backslashbox[10mm]{zenith angle [$^\circ$]}{reconstructed energy [TeV]}} 
    & 0.1 - 0.3 & 0.3 - 0.5 & 0.5 - 1.0 & 1.0 - 2.0 & 2.0 - 5.0 & 5.0 - 100.0 \\\hline
    0.0 - 15.0 & 120k/240k & 55k/110k & 55k/115k & 35k/70k & 25k/45k & 15k/25k \\
    15.0 - 25.0 & 95k/190k & 60k/120k & 65k/125k & 40k/85k/ & 30k/55k & 15k/35k \\
    25.0 - 35.0 & 60k/120k & 65k/130k & 70k/135k & 50k/95k & 35k/70k & 20k/45k \\
    35.0 - 42.5 & -/- & 75k/150k & 75k/155k & 55k/115k & 45k/95k & 35k/65k \\
    42.5 - 47.5 & -/- & 55k/105k & 95k/195k & 75k/145k & 60k/125k & 50k/100k \\
    47.5 - 52.5 & -/- & -/- & 140k/275k & 100k/200k & 95k/185k & 80k/165k \\
    52.5 - 60.0 & -/- & -/- & 50k/100k & 70k/140k & 70k/135k & 70k/140k \\ \hline
  \end{tabular}
  \caption{Number of signal- (first value) and background training events (second value) in all trained zenith angle- and energy
bands. Events with small energy and large zenith angle cannot be reconstructed since the energy threshold of the H.E.S.S. 
array increases with zenith angle.}
  \label{table:training}
\end{center}
\end{table*}

\begin{figure*}
\begin{center}
\begin{minipage}[t]{8cm}
  \includegraphics[scale=0.425]{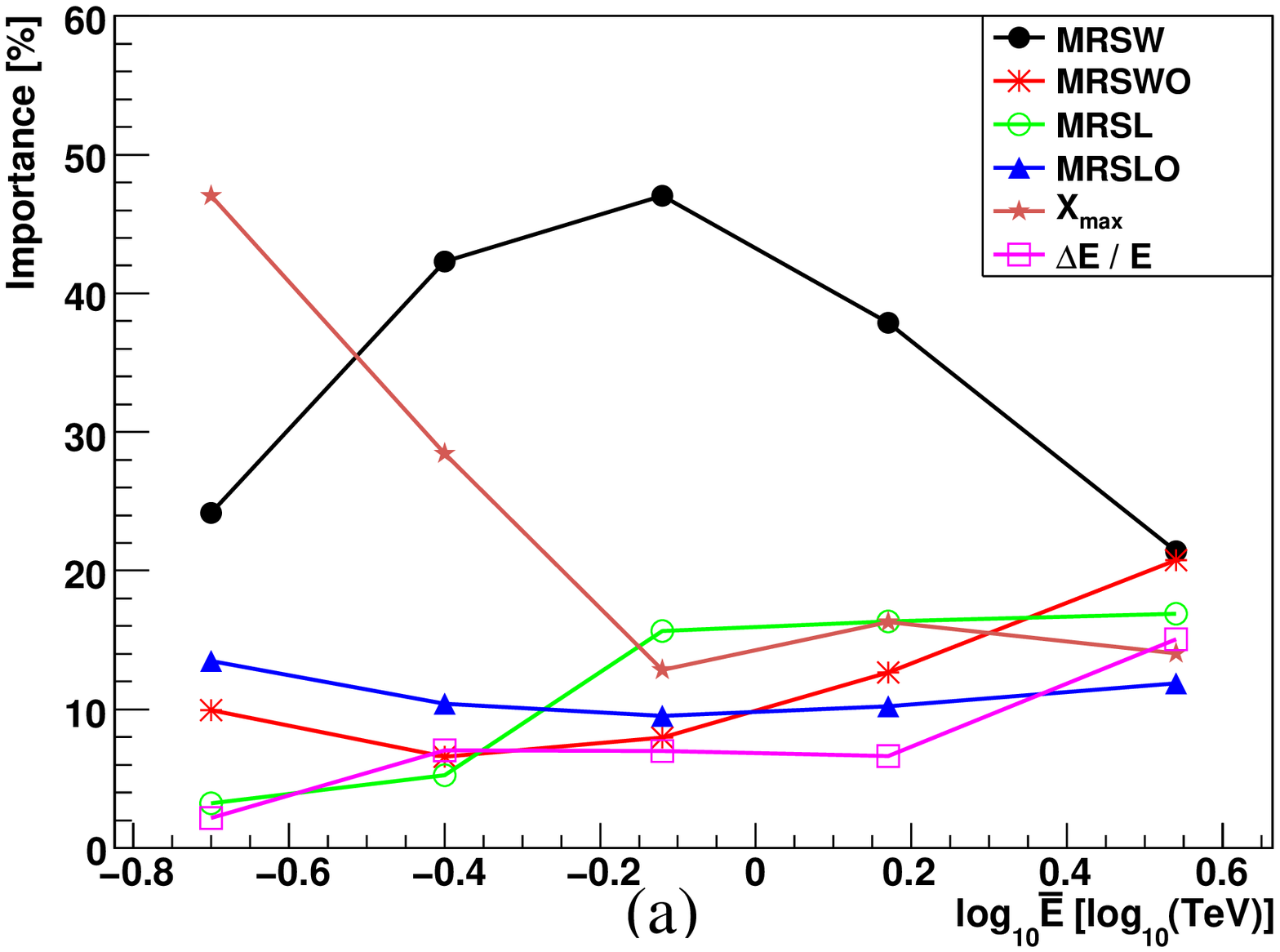}
\end{minipage}
\begin{minipage}[t]{8cm}
  \includegraphics[scale=0.425]{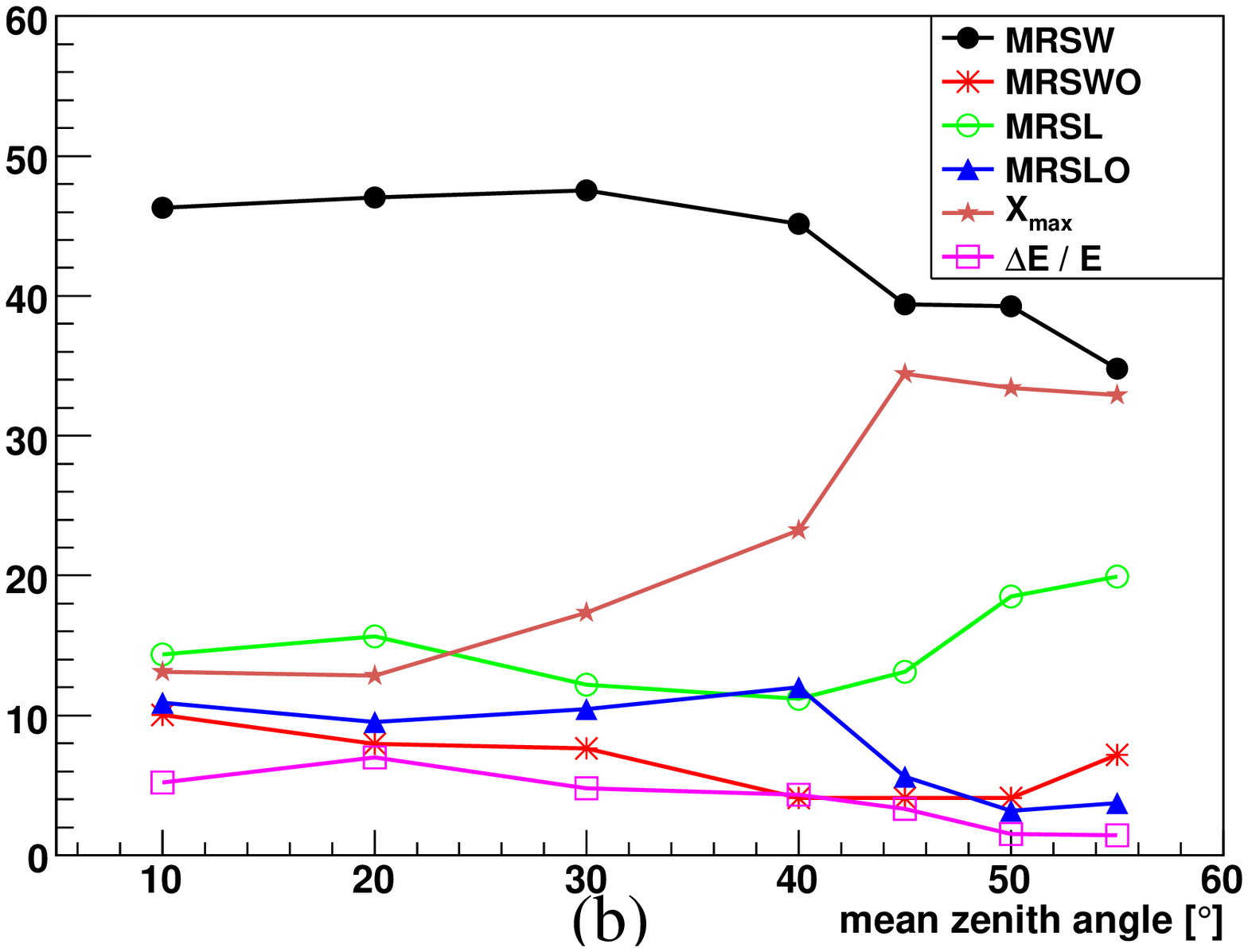}
\end{minipage}
  \caption{Importance (as defined in the main text) of the training parameters as a function of \REMOVE{{\it left}} 
\ADD{{\bf (a)}} mean reconstructed energy in the (15--25)$^{\circ}$ zenith angle band and \REMOVE{{\it right}} \ADD{{\bf (b)}} 
as a function of mean zenith angle for reconstructed energies between (0.5--1.0)~TeV.} 
  \label{fig:importance}
\end{center}
\end{figure*}

As visible from Fig. \ref{fig:trainpar} all parameters show a more or less pronounced separation power which manifests 
itself in a different {\it importance} of these variables for the building of the BDT. This importance is calculated using 
the rate of occurrence of a splitting variable during the training procedure, weighted by the squared separation-gain 
and the number of events in the corresponding nodes \cite{Breiman1984}. Fig. \ref{fig:importance} demonstrates that the 
relative importance of the training parameters does depend on the energy and zenith angle of the event and that this 
importance changes from band to band. 

While the MRSW parameter is generally the most important classification variable, this is not true for events with energies 
below a few hundred GeV, since in this energy range hadron- and $\gamma$-initiated showers look similar \cite{Sobo2007,
Maier2007}. Here, the X$_{\mathrm{max}}$ parameter provides better separation, because it carries information about the 
primary particle interaction length without taking into account the shape of the shower image. Therefore, X$_{\mathrm{max}}$ 
is uncorrelated with the image shape parameters and an important parameter for the $\gamma$/hadron separation at low 
energies and large zenith angles. 

On the other hand, the spread in event energy reconstruction, $\Delta$E / E, becomes more important for 
events of high energies, because in this energy range $\gamma$-initiated showers exhibit a rather regular shape, whereas 
hadron-initiated showers show large fluctuations and therefore a large spread in the energy reconstructed by the 
participating telescopes.

Also the MRSWO and MRSLO parameters carry additional information about the shower shape. They suffer from the larger 
hadronic shower fluctuations, but nevertheless contribute to a significant extent to the training procedure. 

Beyond the parameters used in this work, additional variables which parametrise the intrinsic image properties 
(e.g. like those obtained for the 3D Model analysis \cite{Marianne2006}) are sensitive to different shower properties 
and could further improve the BDT classification.

\subsection{BDT response} \label{subsection:bdtresponse}
\begin{figure}
\begin{center}
  \includegraphics[width=9cm]{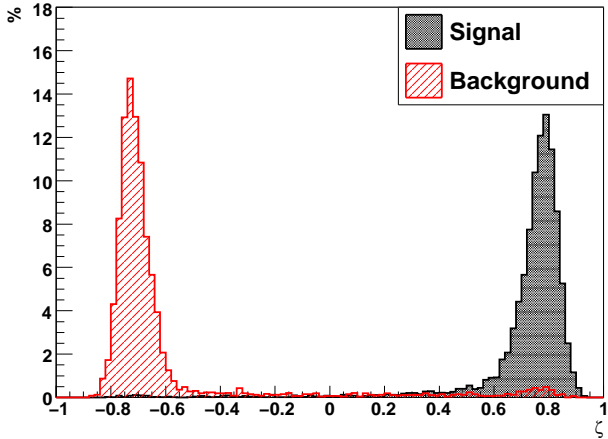}
  \caption{BDT output for events using an independent test sample (same energy and zenith range as in Fig. 
\ref{fig:trainpar}).}
  \label{fig:tmva_output}
\end{center}
\end{figure}

After having grown the BDT, the classifier's response was tested in all zenith angle- and energy bands with an independent 
test sample of signal- and background events. As an example, Fig. \ref{fig:tmva_output} shows the result of the 
classification of this test sample with the BDT trained in the (0.5--1.0)~TeV band with zenith angles (15--25)$^\circ$, 
demonstrating the excellent classification power of the BDT approach in terms of $\gamma$/hadron separation. However, 
as explained in the last section, \ADD{some of the} input parameters depend on zenith angle and energy and therefore the 
$\zeta$ distributions look different from band to band. This later requires zenith- and energy-dependent cuts, to make the 
$\gamma$/hadron separation independent of the input parameter distributions (Chapter \ref{subsection:spectralanalysis}).

\section{Systematic studies using H.E.S.S.  data} \label{section:comparison}

The consistency between data and simulations is one of the key aspects for the analysis of VHE $\gamma$-ray sources. 
Since observations cover a broad energy range and are performed under various observational conditions (e.g. different 
zenith angles or telescope configurations), the BDT classification has to be tested under these conditions. 
For this purpose, we apply the BDT method to H.E.S.S. observations of the Galactic Centre (GC) region 
performed in 2004 and compare the excess of $\gamma$-rays above the background with the predictions from $\gamma$-ray 
simulations with similar properties. 

\subsection{Comparison between simulations and data} \label{section:comparisonmcreal}

The data set used here is a subset of the GC observations \cite{DM2006} and accumulates to a total livetime 
\footnote{The livetime is the observation time corrected for the dead-time of the system.} of 11.4 hours. The data 
were selected by zenith angle 
to cover a smaller range of $15^\circ\leq\theta\leq25^\circ$, thereby avoiding the mixing of $\gamma$-ray simulations 
at different zenith angles when comparing the results. The mean offset of the observations is 1$^\circ$. In the following 
we compare the $\gamma$-ray excess of the GC source HESS~J1745--290 to $\gamma$-ray simulations at a fixed zenith- and offset 
angle of $20^\circ$ and 1$^\circ$, respectively. The energy spectrum of HESS~J1745--290 follows a power-law in energy with 
a spectral index of $\Gamma$=2.21 between (0.2--10.0)~TeV \cite{Aharonian2004}, and the $\gamma$-ray simulations are 
chosen to match the same spectral shape in this energy range.

\begin{figure*}
\begin{center}
  \includegraphics[width=15cm]{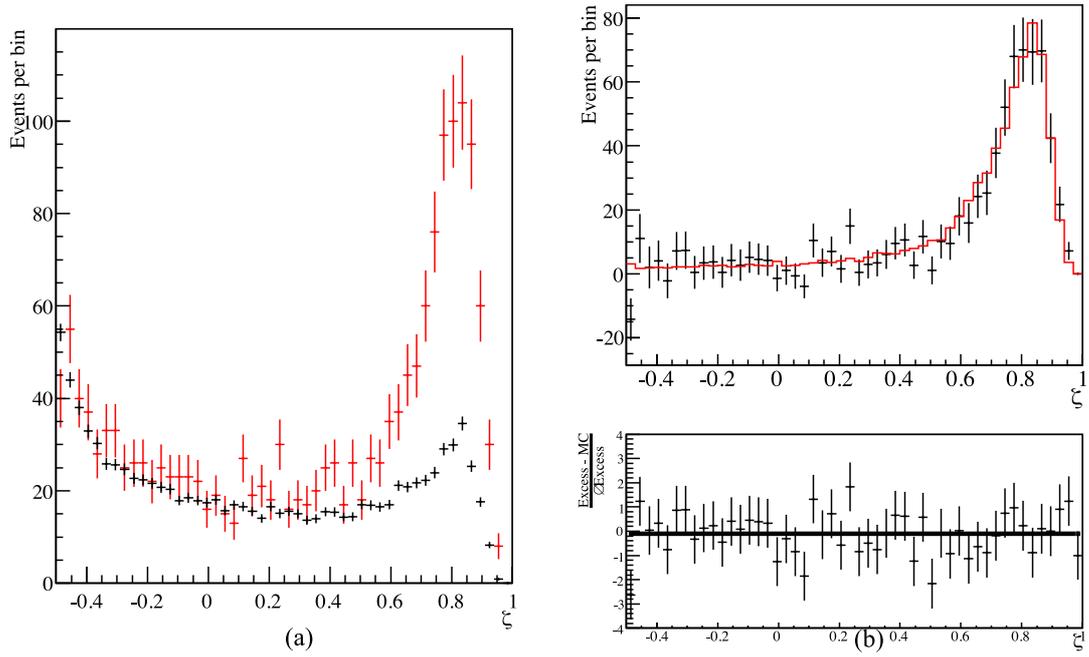}
  \caption{\REMOVE{{\it left}} \ADD{{\bf (a)}}: $\zeta$ distribution for events from the On-Region (red) and events 
from the Off-Regions (black), weighted by $\alpha$, from HESS~J1745--290 observations. \REMOVE{{\it right}} 
\ADD{{\bf (b)}}: Comparison between $\gamma$-ray simulations (red curve) and $\gamma$-ray excess, normalised to the number 
of events in the range ($0 \leq \zeta \leq 1$). Also shown are the residua between the two distributions and the result 
of a fit of a constant, which is compatible with 0 residuum within the statistical errors and has a $\chi^2$/ndf of 40/49.}
  \label{fig:tmvasgr}
\end{center}
\end{figure*}

The $\zeta$ distributions for events coming from the assumed source region (On-Region) and from seven background 
control regions (Off-Regions) \footnote{The used background estimation method is known as reflected background model 
\cite{Berge2007}.} are shown in Fig. \ref{fig:tmvasgr}, \REMOVE{left} \ADD{(a)}. The $\gamma$-ray excess can then be 
calculated as $\mathrm{N_\gamma} = \mathrm{N_{On}} - \alpha\cdot\mathrm{N_{Off}}$, with $\mathrm{N_{On}}$ and $\mathrm{N_{Off}}$ 
being the number of events from the On-Region and Off-Regions, respectively, and $\alpha$ as normalisation factor which 
accounts for the different geometrical areas of the On-Region and Off-Regions. The comparison between $\gamma$-ray excess and 
simulated $\gamma$-rays (Fig. \ref{fig:tmvasgr}, (b)) reveals an excellent agreement and demonstrates that the BDT 
classifies both type of events in the same way in a broader zenith angle- and energy range.

\begin{figure*}
\begin{center}
\begin{minipage}[t]{8cm}
  \includegraphics[scale=0.4]{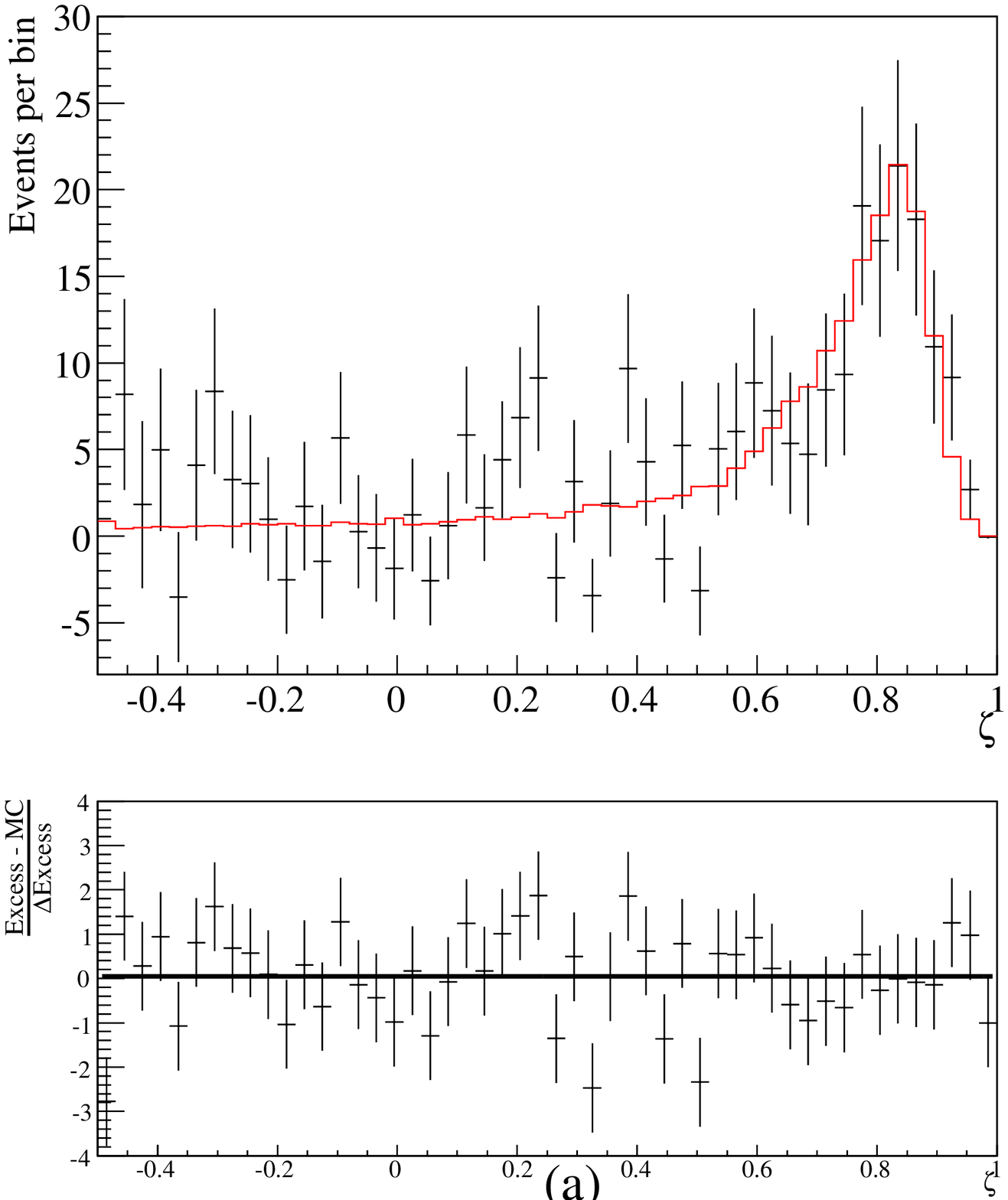}
\end{minipage}  
\begin{minipage}[t]{8cm}
  \includegraphics[scale=0.4]{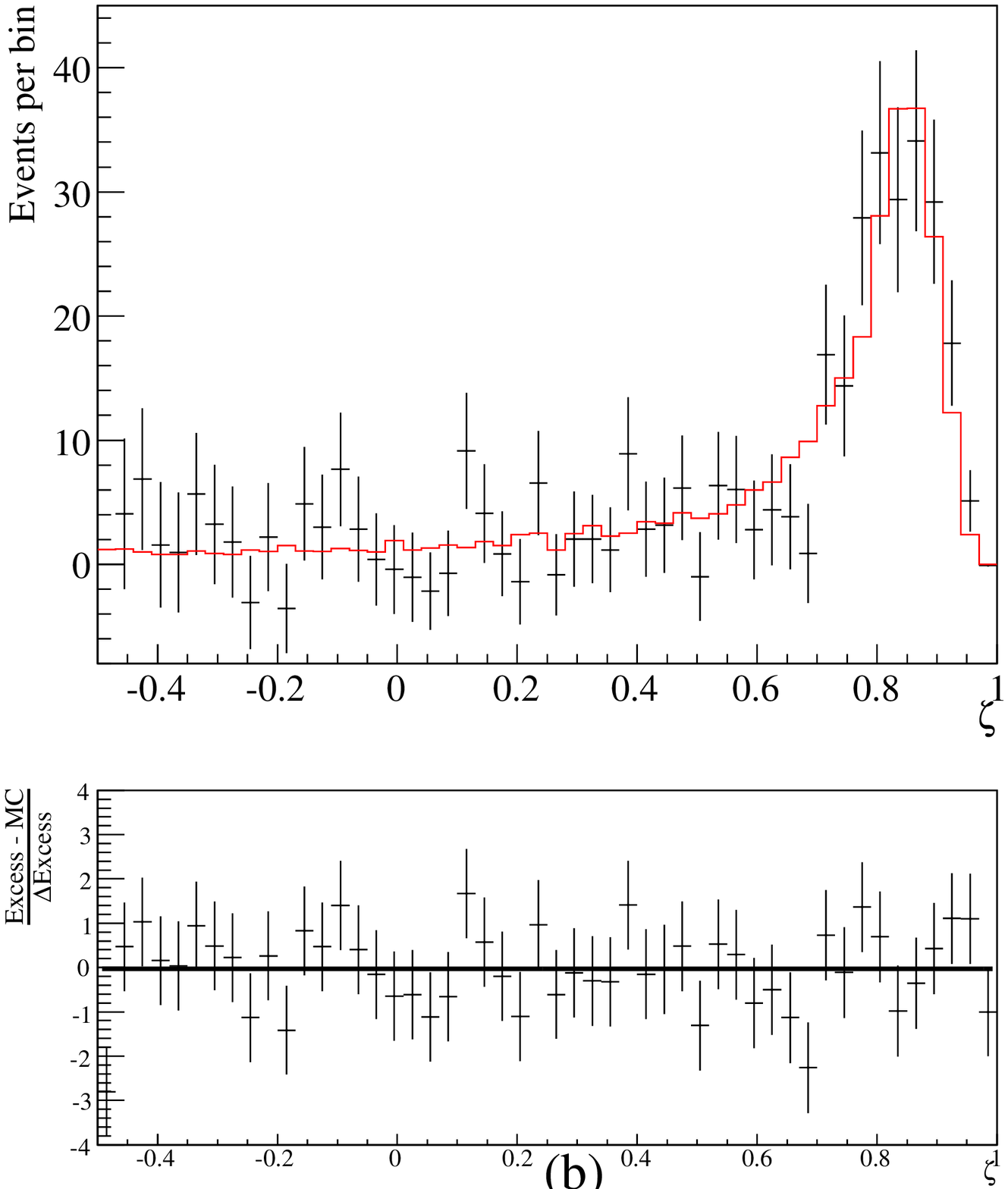}
\end{minipage}  
\caption{Comparison of $\zeta$ distributions for $\gamma$-ray simulations and $\gamma$-ray excess \REMOVE{{\it left}} 
\ADD{{\bf (a)}}: for events with a multiplicity of 2 and \REMOVE{{\it right}} \ADD{{\bf (b)}}: for events with 
reconstructed energies 0.2~TeV $\leq$ E $\leq$ 0.4~TeV. The lower panel again shows their residua and the result of a 
fit of a constant. Both fits are compatible with 0 residuum within the statistical errors and have a $\chi^2$/ndf of 
57/49 and 43/49, respectively.}
  \label{fig:tmvasystematic}
\end{center}
\end{figure*}

To illustrate the stability of the BDT classification with respect to different subsets of events, Fig. 
\ref{fig:tmvasystematic} shows the comparison for events with low energies of 0.2 TeV $\leq$ E $\leq$ 0.4 TeV and for 
events which were recorded by just two telescopes. \ADD{These two subsets contain 1/2 and 1/3 of all events, respectively, 
and are difficult to classify, given the limited separational information for such kind of events. Even for those, the 
agreement between $\gamma$-ray simulations and $\gamma$-ray excess is obvious and confirms the robustness of the BDT 
classification.} \REMOVE{under different observational conditions.}

\subsection{Spectral analysis with the BDT method} \label{subsection:spectralanalysis}

The last section illustrated that the BDT classification of data and simulations leads to consistent results under 
variation of different parameters like the covered energy range or the telescope multiplicity of the events. Hence, 
the BDT classification can be used to select $\gamma$-ray-like events for the spectral analysis of VHE $\gamma$-ray 
sources.

As aforementioned, the energy- and zenith-dependence of \ADD{some of} the input parameters leads to a zenith- and 
energy dependent BDT classification. A fixed cut on $\zeta$ would accordingly lead to different cut efficiencies and 
hence result in a classification which depends on the observational conditions \footnote{\ADD{On the other hand, cuts 
on MRSW and MRSL as applied in the H.E.S.S. Standard Analysis neither depend on the event energy nor on the zenith angle 
and hence preserve the cut efficiency.}}. To circumvent this problem, the independent test sample was used to predict the 
$\gamma$-efficiency of all possible $\zeta$ cuts in each zenith angle- and energy band. This information was then used 
to assign a corresponding $\gamma$-efficiency to every $\zeta$ of an event, $\epsilon_{\gamma}(\zeta)$.

In the H.E.S.S. Standard Analysis \cite{Crab2006}, $\gamma$-ray selection cuts are optimised on MRSW, MRSL, image 
intensity and $\theta^2$ \footnote{The squared angular distance between the assumed source position and the reconstructed 
shower direction.} simultaneously to maximise the significance ($\sigma$, defined in \cite{LiMa1983}, Equation (17)). 
The same optimisation procedure was applied to our analysis, but using $\epsilon_\gamma(\zeta)$ instead of MRSW and MRSL.  
Here we optimised for two different sets of assumed strength and spectral index of the source, namely the $\zeta$ std-cuts 
(10\% \ADD{of the integrated} Crab flux \ADD{above 200~GeV} \REMOVE{and }\ADD{with} a spectral index of $\Gamma$=2.6) 
and the $\zeta$ hard-cuts (1\% \ADD{of the integrated} Crab flux \ADD{above 200~GeV} \REMOVE{and }
\ADD{with} a spectral index of $\Gamma$=2.0). Together with the cuts used in the H.E.S.S. Standard Analysis, which are optimised 
for the same source types, they are summarised and described in Table \ref{table:cuts}.

\begin{table}
\ADD{{\bf (a)}}
  \begin{center}
    \begin{tabular}{|c|c|c|c|} \hline
      Configuration & $\epsilon_{\gamma}(\zeta)$ & $\theta^2_{cut}$ & Size \\ 
       & Max & Max & Min \\
       & & (degrees$^2$) & (p.e.) \\ \hline
      Standard &  0.84 & 0.0125 & 60 \\
      Hard &  0.83 & 0.01 & 160 \\ \hline
    \end{tabular}\\
  \end{center}
\ADD{{\bf(b)}}
  \begin{center}
    \begin{tabular}[]{|c|cccc|} \hline
      & MRSW & MRSL & $\theta^2_{cut}$ & Size \\
      Configuration & Max & Max & Max & Min \\
      & $\sigma$ & $\sigma$ & (deg$^2$) & (p.e.) \\ \hline
      Standard & 0.9 & 2.0 & 0.0125 & 80 \\ 
      Hard & 0.7 & 2.0 & 0.01 & 200 \\ \hline
    \end{tabular}
    \caption{\REMOVE{{\it top}} \ADD{{\bf (a)}}: Selection cuts optimised for Configuration Standard (strong, steep spectrum 
      sources) and Hard (weak, hard spectrum sources) for the $\zeta$ analysis. \REMOVE{{\it bottom}} \ADD{{\bf (b)}}: 
      Selection cuts optimised for the Standard and Hard Configurations for the H.E.S.S. Standard Analysis \cite{Crab2006}. 
      Minimum cuts on MRSW and MRSL of -2.0 are applied in the case of the Standard Analysis.}
    \label{table:cuts}
  \end{center}
\end{table}

\begin{table*}
\begin{center}
  \begin{tabular}{|c|c|c|c|c|c|c|} \hline
    \multicolumn{1}{|c|}{\backslashbox[10mm]{zenith angle [$^\circ$]}{reconstructed energy [TeV]}} 
    & 0.1 - 0.3 & 0.3 - 0.5 & 0.5 - 1.0 & 1.0 - 2.0 & 2.0 - 5.0 & 5.0 - 100.0 \\\hline
    0.0 - 15.0 & 0.28/0.31 & 0.59/0.61 & 0.63/0.64 & 0.52/0.59 & 0.61/0.62 & 0.63/0.64 \\
    15.0 - 25.0 & 0.27/0.29 & 0.56/0.58 & 0.61/0.63 & 0.56/0.57 & 0.56/0.57 & 0.60/0.61 \\
    25.0 - 35.0 & 0.22/0.25 & 0.51/0.53 & 0.59/0.60 & 0.55/0.57 & 0.48/0.51 & 0.54/0.56 \\
    35.0 - 42.5 & -/- & 0.45/0.48 & 0.58/0.60 & 0.52/0.53 & 0.44/0.46 & 0.43/0.45 \\
    42.5 - 47.5 & -/- & 0.25/0.28 & 0.54/0.56 & 0.54/0.56 & 0.42/0.45 & 0.39/0.42 \\
    47.5 - 52.5 & -/- & -/- & 0.47/0.50 & 0.48/0.51 & 0.36/0.39 & 0.38/0.41 \\
    52.5 - 60.0 & -/- & -/- & 0.29/0.32 & 0.46/0.48 & 0.38/0.40 & 0.35/0.37 \\ \hline
  \end{tabular}
  \caption{$\zeta$ cuts in all zenith angle and energy bands which correspond to an $\epsilon_{\gamma}(\zeta)$ cut of 0.84 
($\zeta$ std-cuts, first value) and 0.83 ($\zeta$ hard-cuts, second value).}
  \label{table:effzeta}
\end{center}
\end{table*}

\begin{figure}
\begin{center}
  \includegraphics[width=9cm]{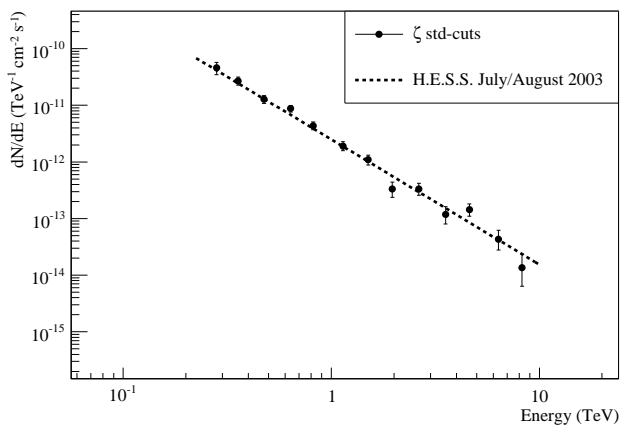}
  \caption{Comparison between the fit to the energy spectrum of HESS~J1745--290 for the July/August 2003 data set (dashed 
line, \cite{Aharonian2004}) and the spectrum obtained with the $\zeta$ std-cuts (filled circles).}
  \label{fig:spectrum}
\end{center}
\end{figure}

The optimised $\zeta$ std-cuts were applied to the HESS~J1745--290 data set and a spectrum was extracted. The 
spectrum obtained for the application of the $\zeta$ std-cuts and the published differential flux \cite{Aharonian2004} 
are shown in Fig. \ref{fig:spectrum}. An excellent agreement between both results further consolidates the applicability of 
the BDT approach for the analysis of VHE $\gamma$-ray sources. Additional spectral tests with sources of different spectral 
shape, flux level or source extensions were performed. They show the same agreement between the $\zeta$ std-cuts and 
the H.E.S.S. Standard Analysis.

\section{Performance and Sensitivity}  \label{section:tests}

Having shown the applicability of the BDT classification under different observational conditions and for the spectral 
analysis, the performance and sensitivity of BDT is studied on the basis of $\gamma$-ray simulations and Off-Events.

\subsection{Separation power of $\zeta$ cuts}
\begin{figure*}
  \begin{center}
    \begin{minipage}[t]{8cm}
      \includegraphics[scale=0.42]{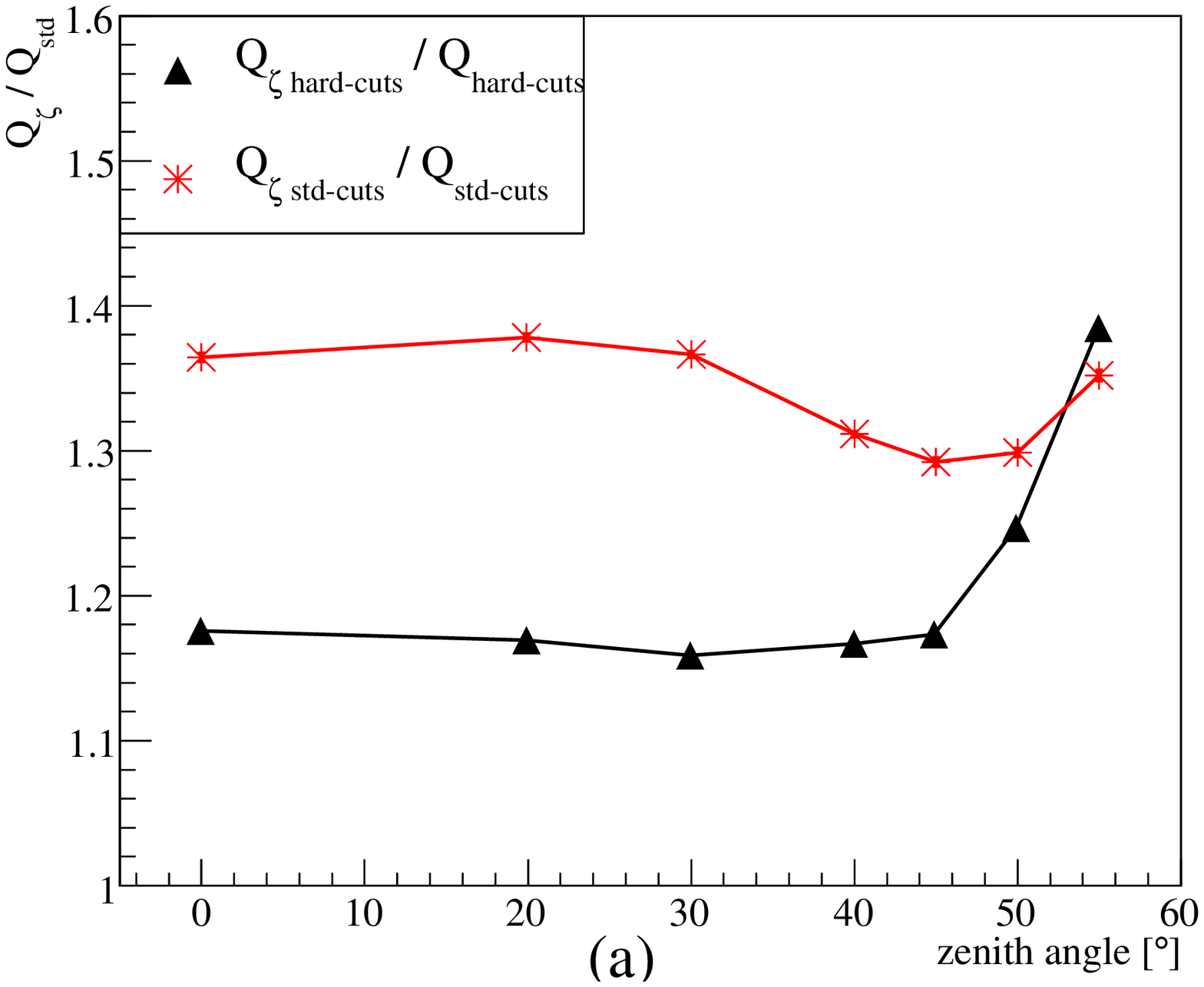}
    \end{minipage}
    \begin{minipage}[t]{8cm}
      \includegraphics[scale=0.42]{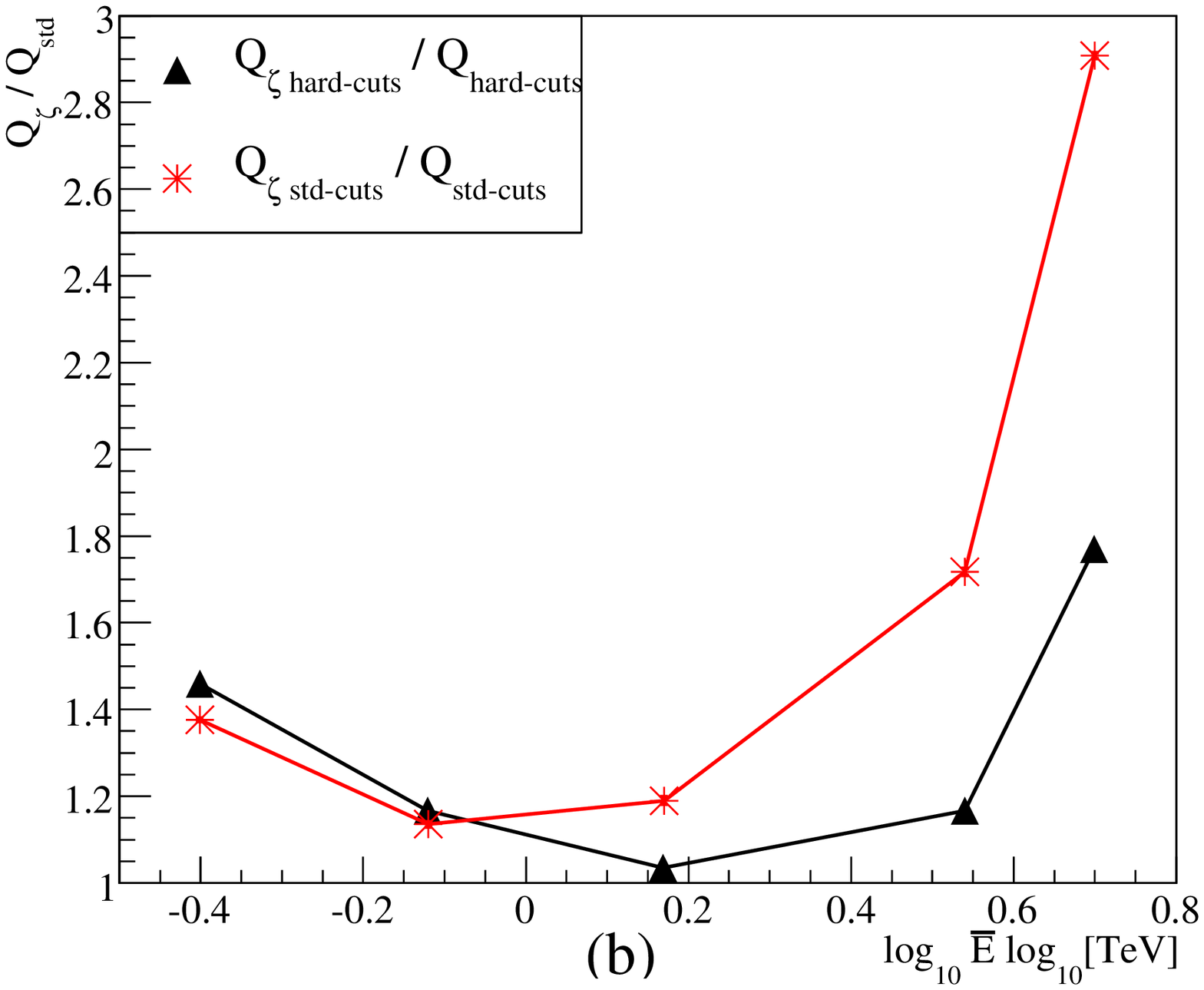}
    \end{minipage}
  \caption{Improvement in Q-factor, defined as Q$_{\zeta}$ / Q$_{std}$ (Q-factor described in the main text) 
versus \REMOVE{{\it left}} \ADD{{\bf (a)}}: zenith angle and \REMOVE{{\it right}} \ADD{{\bf (b)}}: reconstructed energy for the $\zeta$ analysis compared to the H.E.S.S. 
Standard Analysis. The 
values for {\it Q} are determined using the cut efficiencies for $\gamma$-rays from simulations and background 
events from off source data in the specific zenith angle and reconstructed energy range.}
  \label{fig:zenith}
  \end{center}
\end{figure*}

An appropriate parameter to quantify the quality of analysis cuts is the {\it quality factor Q} (e.g. \cite{Bugayov2002}), 
defined as:

$$\mathrm{Q} = \frac{\epsilon_{\gamma}}{\sqrt{\epsilon_{CR}}}\mathrm{,}$$
$$\mathrm{with}\qquad \epsilon_{i} = \frac{\hat{N_i}}{N_i} \qquad(i = \gamma~\mathrm{or~CR}),$$
where the cut efficiency $\epsilon_i$ is defined as the number of events that pass certain cuts $\hat{N_i}$ divided 
by the number of events before cuts $N_i$. 
Fig. \ref{fig:zenith} shows the development of {\it Q$_{\zeta}$ / Q$_{std}$} as a function of zenith angle and energy 
for the $\zeta$ std- and hard-cuts and the std- and hard-cuts after application of the pre-selection and the image shape 
selection. The pre-selection consist of the corresponding image size cuts and a cut on the distance between the $COG$ of 
the shower image and the camera centre to avoid truncated images at the camera edge. The image shape cuts comprehend 
cuts on $\zeta$ and MRSW, MRSL, respectively (see Table \ref{table:cuts} and \ref{table:effzeta} for further information). 

The training in energy- and zenith angle bands leads to a stable improvement in separation power for the BDT 
method and makes the zenith- and energy-dependent cuts on $\zeta$ well suited for this kind of analysis. Especially at energies 
below a few hundred GeV and energies above a few TeV the improvement in Q for hard- and std-spectrum sources is remarkable. 
As a result of the training with $\gamma$-rays simulated at a fixed offset of 0.5$^\circ$, the performance of the $\zeta$ cuts 
is reduced for events with larger offsets ($\geq1.5^\circ$). However, a training in offset bands resulted in steps in selection 
efficiency across the field of view and in the description of the camera acceptance, and is therefore not employed.

\subsection{Sensitivity of $\zeta$ cuts}

\begin{figure*}
\begin{center}
\begin{minipage}[t]{8cm}
  \includegraphics[scale=0.4]{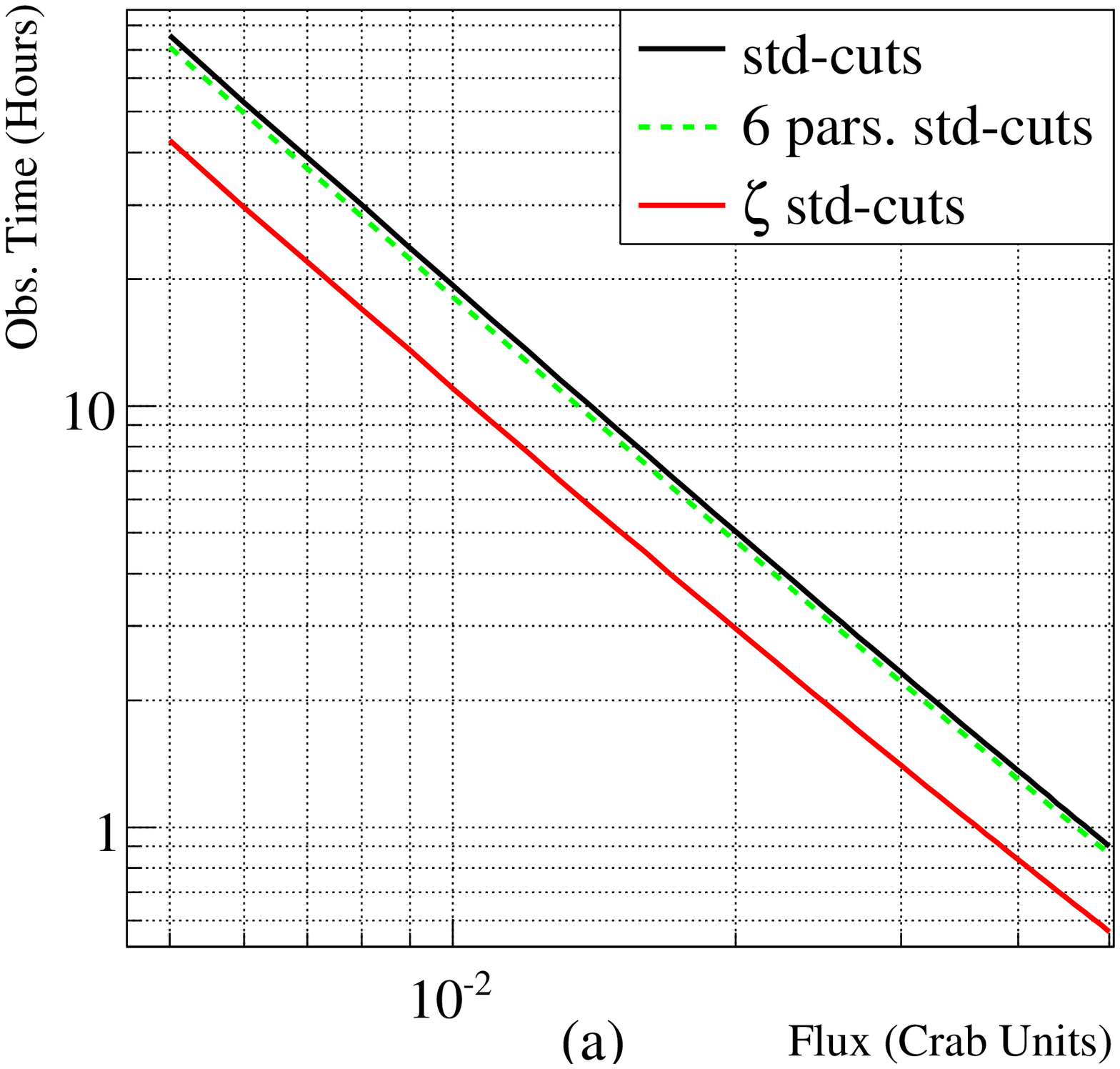}
\end{minipage}
\begin{minipage}[t]{8cm}
  \includegraphics[scale=0.4]{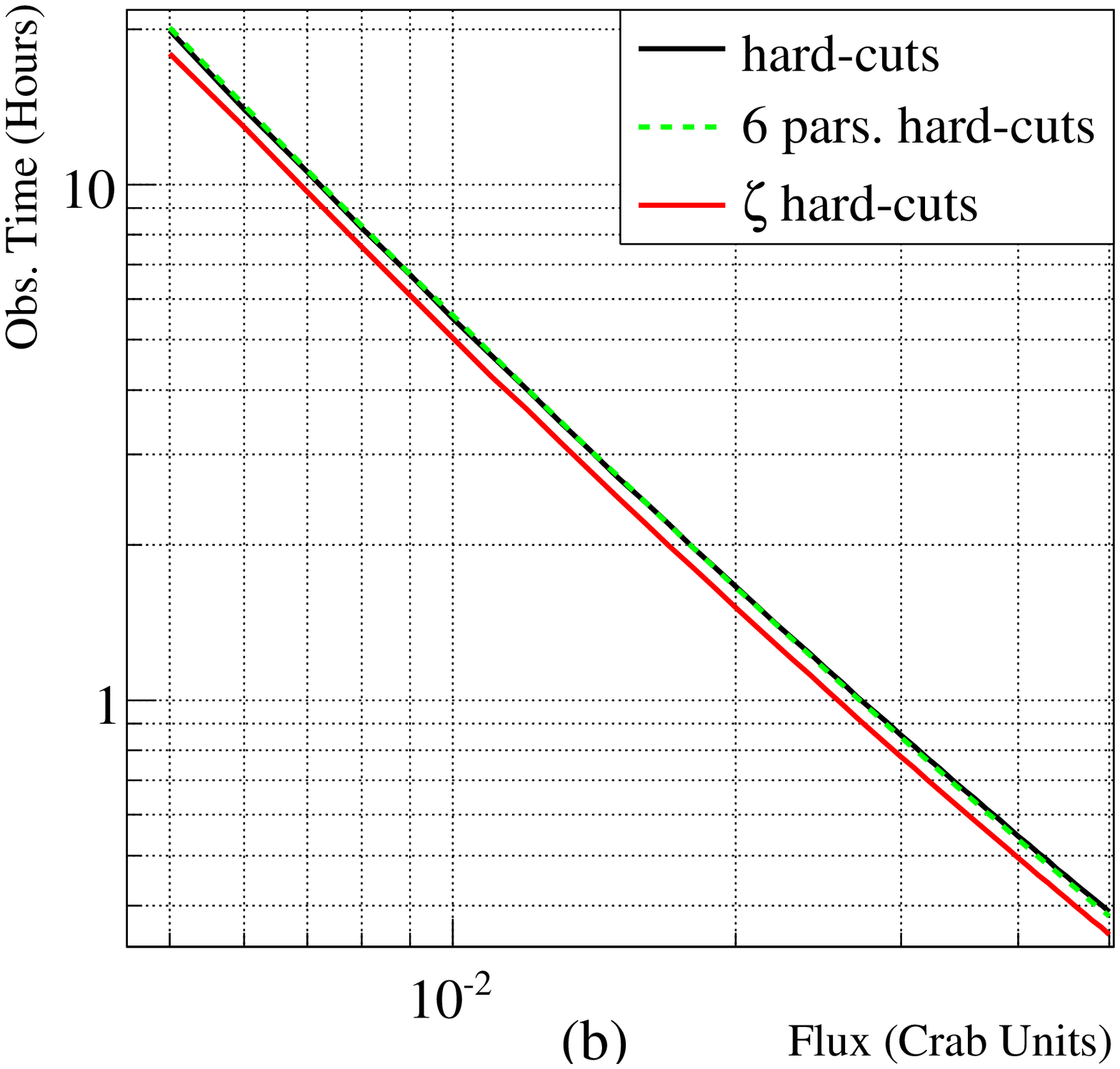}
\end{minipage}
  \caption{Sensitivity of the H.E.S.S. array for six sets of selection cuts. Shown is the required observation time 
to detect a point-like $\gamma$-ray source with 5$\sigma$ significance above the background, as a function of the 
flux of the source with spectral index \REMOVE{{\it left}} \ADD{{\bf (a)}} $\Gamma = 2.63$ and \REMOVE{{\it right}} 
\ADD{{\bf (b)}} $\Gamma = 2.0$ at 20$^{\circ}$ zenith angle and 0.5$^{\circ}$ offset. \ADD{Note, that the required observation 
time is up to 45\% and 20\% less for the BDT method compared to the H.E.S.S. Standard Analysis for configuration Standard and 
Hard, respectively. Furthermore, no gain in sensitivity is achieved in the case of the six parameter box cuts optimisation for 
configuration Hard (black curve is hidden behind the green, dashed curve).}}
  \label{fig:sensitivity}
\end{center}
\end{figure*}

The optimised $\zeta$ cuts (introduced in Section \ref{subsection:spectralanalysis} and Table \ref{table:cuts}) 
are applied to $\gamma$-ray simulations and Off-Events, and their sensitivity for strong, std-spectrum and weak, hard-spectrum 
sources was calculated. \ADD{To disentangle the performance improvement due to the information stored in the additional 
parameters and due to the treatment of non-linear correlations by the BDT method, the sensitivity for optimised 
{\it box cuts} on all training parameters is also shown. These box cuts are a set of one-dimensional cuts on each 
training parameter which are all optimised simultaneously to obtain the best separation between signal and background.}

Fig. \ref{fig:sensitivity} illustrates the improved separation power of the $\zeta$ analysis compared to the 
\REMOVE{{\it box cuts}} \ADD{box cuts} applied in the H.E.S.S. Standard Analysis. Shown is the required observation time for a 
detection (signal with more than 5$\sigma$ above background) of a point source for a range of fluxes, assuming a power-law 
in energy with a spectral index of $\Gamma = 2.63$ as measured for the Crab nebula \cite{Crab2006} (Fig. 
\ref{fig:sensitivity}, \REMOVE{left} \ADD{(a)}) and for a hard spectrum source with index $\Gamma = 2.0$ (Fig. 
\ref{fig:sensitivity}, \REMOVE{right} \ADD{(b)}) for the aforementioned sets of selection cuts. Remarkably, the optimised 
$\zeta$ cuts show the highest sensitivity over a wide range of source strengths. The required observation time for the 
$\zeta$ analysis is up to ~45\% and ~20\% less compared to the H.E.S.S. Standard Analysis, for configuration Standard and 
Hard, respectively. \ADD{It is also clear from Fig. \ref{fig:sensitivity}, that box cuts add just little to the total 
separation gain, since they ignore non-linear correlations in the six training parameters.}

Since the X$_{\mathrm{max}}$ parameter contributes especially at low energies to the classification (see Fig. 
\ref{fig:importance} for comparison), and cuts optimised for hard-spectrum sources tend to reject low-energy events, the 
performance improvement of the $\zeta$ hard-cuts is only 20\% compared to the H.E.S.S. hard-cuts. Nevertheless, the 
improvement is stable over a wide range of fluxes. One possibility to further improve the BDT performance for hard-spectrum 
sources is to find the best match between size cut applied to select the training sample (see Section 
\ref{subsection:trainingsample} for comparison) and size cut optimised for a given source type in an alternating process. 

\section{Summary and Outlook}

IACTs have to deal with a vast number of hadronic cosmic-ray background events. The capability to suppress these against the 
$\gamma$-rays is one of the aspects which limits the sensitivity of IACTS. In this work the training, testing and evaluation 
of the BDT method with H.E.S.S. data was presented. The BDT is a multivariate analysis method, which combines the information 
carried in several classification parameters into one parameter $\zeta$. This parameter describes the likeness of 
an event to be of hadronic or electromagnetic origin. Observations of the VHE $\gamma$-ray source HESS~J1745--290 
show a very good agreement between $\zeta$ distribution of the measured $\gamma$-ray excess and the predictions from 
$\gamma$-ray simulations for a variety of observational conditions. Zenith- and energy-dependent cuts are introduced to 
account for the zenith and energy-dependent classification of the BDT. Performance tests have shown a dramatically 
increased separation power for the $\zeta$ analysis compared to the H.E.S.S. Standard Analysis especially for sources with 
a spectral index compatible with that measured for the Crab nebula.

The systematic studies performed in this work and the achieved classification power demonstrate that a multivariate analysis 
approach like BDT is well suited for the analysis of $\gamma$-ray data measured with instruments like H.E.S.S.. In near- and 
mid-term projects like H.E.S.S. II, MAGIC II, CTA and AGIS the accessible energy range of IACTs is extended as the reachable 
sensitivity increases. Multivariate methods can play a major role for the analysis and particularly for the $\gamma$/hadron 
separation of upcoming instruments. The majority of the events will be recorded below a 100~GeV, where $\gamma$/hadron 
separation is increasingly difficult. In this work, performance of parameters such as X$_{\mathrm{max}}$ demonstrate the
ability especially for the separation at low energies.

\section*{Acknowledgement}

The authors would like to acknowledge the support of their host institution, and additionally support from the German 
Ministry of Education and Research (BMBF). We would like to thank the whole H.E.S.S. collaboration for their support, 
especially Werner Hofmann and our H.E.S.S. colleagues from the APC, Paris and the University of Erlangen-N\"urnberg for 
fruitful discussions.


\end{document}